\documentstyle[12pt]{article}
\begin{document}

\newcommand {\eqn}[1]{\begin{equation}#1\end{equation}}
\newcommand{\equln}[2]{\begin{equation} {#1} \label{#2}\end{equation}}
\newcommand{\includegraphics}[1]{\zscale{100}\hbox{\epsffile{#1}}}

\title{\large Fundamental excitations in layered superconductors with long-range
Josephson couplings}
\author {by\\Ma{\l}gorzata Sztyren
\thanks{Department of Mathematics and Information Science,
Warsaw University of Technology, Pl. Politechniki 1, PL-00-661 Warsaw
E--mail: emes@mech.pw.edu.pl}}

\input{epsf.tex}
\input{trans.tex}
\special{ps:}

\maketitle

\begin{abstract}
  The present paper develops the ideas introduced in {\em cond-mat/0312673}.
The construction of a hybrid discrete-continuous model of layered
superconductors is briefly presented. The model bases on the classic
Lawrence-Doniach scenario with admitting, however, long-range
interactions between atomic planes. Moreover, apart from Josephson
couplings they involve the proximity effects. The range of
interactions, K, can, in principle, be arbitrary large. The solutions
corresponding to the range K=2 are exposed.
  The fundamental excitations are understood as deviations from stable
ground states.The formulae for energy of those excitations are constructed.
The possible shapes of dispersion curves are analysed. For each type
of shape the corresponding values of physically measurable quantities
like effective mass and bandwidth are expressed by coupling parameters. 
\end{abstract}                  

\section{Introduction}
The layered structure of superconductor implies a very strong
structural anisotropy which may be characterised by relation between
 \(\xi_c\) --
the coherence length in the direction perpendicular to layers
-- and the interlayer distance \(s\). In the limiting case
\(\xi_c >> s\) the 3D anisotropic Ginzburg--Landau model is well
applicable
\cite{Abrikosov:book,Blatter+:94,Rogula:99,Ketterson+Song}.
On the other side, in the case \(\xi_c\leq s\), a much better
description is given by Lawrence--Doniach model: a stack of
2D GL planes with Josephson links between adjacent planes
\cite{Lawrence+Doniach:71,Kuple:2001}. Similar model has
been used in \cite{Theo:88,Theo:93,Laza+:93,Krasnov:2001}.
However, there exist superconducting materials like
\({\em\ Bi_2Sr_2CaCu_2O_8}\) or
\({\em Tl_2Ba_2CaCu_2O_8}\), which
 have \(\xi_c/s\) between 1.5 and 2, i.e. too small for GL and to
high for LD.\\

   The higher grade hybrid model (HM) \cite{Sztyren:2003,
Rogula+Sztyren:2006,Sztyren:2006,Sztyren:Itaka2006}, presented here,
patches the gap between those extreme cases, admitting
Josephson-type couplings (we shall call them J-links) between
distant planes. The range K of J-links is a global parameter of the model.
K=0 denotes the stack of non-interacting GL planes,
K=1 -- J-links between nearest planes only,
K=2 -- J-links between nearest and next nearest planes,
and so on.\\

   Like in LD model we consider two continuous independent variables
understood as in-plane coordinates \(x, y\). The third independent
variable
is discrete: it denotes the ordinal number of the plane \(n\).\\

   We shall use the following notation:\\
\(\left\{\begin{array}{l}
\left.\begin{array}{l}
\psi_n - \mbox{the order parameter associated to the layer indexed by n}\\
\bar{\psi}_n - \mbox{its complex conjugate (c.c.),}
\end{array}\right.\\
\left.\begin{array}{ll}
m_{ab} & \mbox{- in-plane}\\
m_c & \mbox{out-of-plane}
\end{array}\right\} \mbox{effective mass of superconducting current carriers.}\\
\end{array}\right.\)
\section{Free energy functional}
Contribution to the superconducting component of the free energy contains
2 parts: from GL planes and from J-links.
\begin{eqnarray}
{\cal{F}}_s= s\sum_n\int dxdy\{\frac{\hbar^2}{2m_{ab}}|({\bf D}\psi)_n|^2 
 +\alpha_0|\psi_n|^2+\frac{1}{2}\beta|\psi_n|^4\}\nonumber\\
+\frac{s}{2}\sum_{n\epsilon P}\sum _{q\epsilon P_n}
\int dxdy\,\{\zeta_q(|\psi_n|^2+|\psi_{n+q}|^2)
-\gamma_q
 (\bar{\psi}_n\psi_{n+q}e^{-ip_{qn}}+c.c.)\}
,\label{fj}
\end{eqnarray}
where \(P\) denotes the set of indices of all planes,
\(P_n\) refers to planes
which are J-linked to the plane \(n\). Every J-link is represented by
exactly one term.
The symbol {\bf D} denotes the 2-dimensional operator of covariant
derivative
\begin{equation}
D_{\rho}=\partial _{\rho}-\frac{ie^*}{\hbar c}A_{\rho},
 \ \ \ \rho=x,y.
\end{equation}
The third component of the vector potential \(A_z\) apears in the exponent
\begin{equation}
p_{qn}=\frac{e^*}{\hbar c}\int_{ns}^{(n+q)s}A_zdz.\label{pqn}
\end{equation}
%
%
\section{Comparison with the anisotropic GL model}\label{compa}
To compare our hybrid model (HM) with the continuum GL model,
we shall consider the infinite medium. In that case the component
of free energy connected with J-links has the form
\begin{equation}
F_J=\frac{1}{2}\sum_n\sum_q\{2(\zeta_q-\gamma_q)|\psi_n|^2
 +\gamma_q|\psi_{n+q}e^{-ip_{qn}} -\psi_n|^2\}.\label{FJ}
\end{equation}
For very weak fields \(A_z\) and very slow dependence of \(\psi_n\)
on \(n\) we have the correspondence rules which, in the long wave
limit, allow us
to pass from the hybrid to the 3D continuum.
Applying the rules formulated in \cite{Sztyren:Itaka2006}
one obtains
%
\begin{equation}
{\cal F}_s \rightarrow \int d^3x\{\frac{\hbar^2}{2m_{ab}}
 |{\bf D}\psi|^2+\frac{1}{2}q^2s^2\gamma_q|D_3\psi|^2
 +\alpha_0|\psi|^2
 +\sum_q[(\zeta_q-\gamma_q)|\psi|^2+\frac{1}{2}\beta|\psi|^4]\}
.
\end{equation}
%
It implies the modification of GL parameter \(\alpha_0\) 
to the form
\equln{\alpha=\alpha_0+\sum_q(\zeta_q-\gamma_q)
}{alfa}
and gives the
following relation between the GL effective
mass in \(z-\)direction and the coupling
parameters.
\begin{equation}
m_c=(\frac{s^2}{\hbar^2}\sum_qq^2\gamma_q)^{-1},\ \ \ 
\label{mass}
\end{equation}
%

\section{Field equations}\label{field}
By computing the variation of the functional
 \({\cal F}_s\) with
respect to \(\bar{\psi}_n\), one obtains the equations

\begin{equation}
  -\frac{\hbar^2}{2m_{ab}}{\bf D}^2\psi_n+\tilde{\alpha}\psi_n+
     \beta|\psi_n|^2\psi_n
    -\frac{1}{2}\sum_{q\epsilon \bar{Q}}
    \gamma_q
    \sigma_{qn}\psi_{n+q}e^{-ip_{qn}}=0
\label{eqpsi}
\end{equation}
where
%
\begin{equation}
\tilde{\alpha}=\alpha_0+\frac{1}{2}
      \sum_{q\,\epsilon \bar{Q}}\sigma_{qn}\zeta_q\label{tilal}
\end{equation}
 depends of \(n\) for finite \(P\). The quantities
\begin{equation}
\bar{Q}=\{-K,...,-2,-1,1,2,...,K\},\ \ \ \sigma_{qn}=\left\{ \begin{array}{l}
1\ \ {\rm if}\ \ \ (n+q)\ \epsilon P
\\
0\ \ {\rm otherwise}
\end{array}
\right .\label{sig}
\label{qbar}
\end{equation}
have been introduced to describe the finite range of J-links.\\

   The expression for Josephson current \(J_l\) describing tunelling
across the interplanar gap indexed by \(l\) (half integer if planes
are indexed by integers) will have the form
\begin{equation}
J_l=\frac{se^*}{2i\hbar}\sum_{q\epsilon \bar{Q}}
\sum_{n\epsilon P_{lq}}
 \{\gamma_q\sigma_{qn}\bar{\psi}_n\psi_{n+q}e^{-ip_{qn}}-c.c.\},
 \label{cnb}
\end{equation}
where
\begin{equation}
P_{lq}=\{n\,\epsilon\,P: n<l<n+q\}.
\end{equation}
%

\section{The ground states}\label{ground}
Consider now the plane-uniform states of HM in
the absence of magnetic field. Let us concentrate on the case
\(K=2\). The condition of vanishing Josephson current
is equivalent to
\begin{equation}
\gamma_1(\bar\psi_n\psi_{n+1}-c.c.)
 +\gamma_2(\bar\psi_n\psi_{n+2}
 +\bar\psi_{n-1}\psi_{n+1}-c.c.)=0,
\end{equation}
and the equations (\ref{eqpsi}) take the form
\begin{equation}
\tilde{\alpha}\psi_n +\beta|\psi_n|^2\psi_n-\frac{1}{2}[\gamma_1
(\psi_{n+1}+\psi_{n-1})+\gamma_2(\psi_{n+2}+\psi_{n-2})]=0.
\label{far}
\end{equation}
Solutions with constant amplitude and difference of phase between adjacent
atomic planes, obtained  by the ansatz \(\psi_n=C\,e^{in\theta}\) form
3 classes of states with \(C^2=-\alpha^*\)/\(\beta\): 
\begin{eqnarray}
uniform: & \ \psi_n=C, & \ \alpha^*=\alpha_0 + \zeta_1 + \zeta_2
-\gamma_1 - \gamma_2,\nonumber\\
alternating: & \ \psi_n=(-1)^n C, & \ \alpha^*=\alpha_0 + \zeta_1 + \zeta_2
+\gamma_1 - \gamma_2,\nonumber\\
phase\ modulated: & \ \psi_n=Ce^{\pm in\theta}, &\ \ 
  \theta={\rm \arccos}(-\frac{\gamma_1}{4\gamma_2}).\label{arccos}
\end{eqnarray}
The phase modulared solutions exist if the parameters
\(\gamma_1\) and \(\gamma_2\) fulfill the relation
\(|\gamma_1|\,{\leq}4|\,\gamma_2|\).
Then \(\alpha^*\) is connected with the coupling
constants by the formula
\begin{equation}
\alpha^*=\alpha_0+\zeta_1+\zeta_2+\gamma_2(1+\frac{\gamma_1^2}
  {8\gamma_2^2}).\label{al3}
\end{equation}

     The straight lines
\(\gamma_1+4\gamma_2=0\),  \(\gamma_1-4\gamma_2=0\), and
\(\gamma_1=0\)
divide the plane (\(\gamma_1,\gamma_2\)) into 3 regions 
(shown in Fig. 1 of \cite{Sztyren:Itaka2006}).
Both the uniform  and the
alternating 
solutions always exist. However, in the region \(\bigcirc\)\hskip-0.9em\(=\)
: \( \gamma_1 >0,\ \gamma_1+4\gamma_2>0\),
only the uniform solution is stable,
while in the region
\(\bigcirc\)\hskip-0.9em\(\pm\)
: \( \gamma_1 <0,\ \gamma_1-4\gamma_2<0\),
only the alternating solution is stable. The region
\(\bigcirc\)\hskip-0.9em\(\approx\)
: \( \gamma_2 <0,\ 4\gamma_2<\gamma_1<-4\gamma_2\),
excludes the stability of both the uniform and the alternating
solutions
but, in contrast to that, ensures the existence and stability of
the phase modulated solutions .
In the sector (N): \( \gamma_2 >0,\ |\gamma_1|<4\gamma_2\),
the phase modulated solutions exist but are unstable.\\

   Note that, irrespectively of the values
and signs of \(\gamma_1,\,\gamma_2\), a stable ground state
solution always exists.\\

   For suitable relations between the coupling constants
\(\gamma_1\) and \(\gamma_2\), one can make the parameter
\(\alpha^*\) more negative than \(\alpha_0\). In consequence,
the 3D superconductivity can turn out enhanced with respect to
the 2D superconductivity in the atomic planes. The presence
of coupling constants \(\zeta_q\) allows to take into account
the proximity effect between atomic planes.
According to the formula (\ref{al3}) one can
obtain a negative \(\alpha^*\) starting from a positive \(\alpha_0\).
It means that under appropriate coupling
one can obtain supercoductivity by stacking planes which originally
are in normal state.
\newpage
  The HM with K=1 has in general two coupling parameters. 
\(\zeta\) and \(\gamma\). In particular cases it is related to known
models:\\
\(\zeta=\gamma>0\ \Rightarrow\) LD model, uniform solution, no enhancement.\\
\(\zeta=\gamma<0\ \Rightarrow\) Theodorakis model, alternating solution,
proximity effect,
\vskip 0.0in\hskip 0.85in enhancement.

\section{The excited states}
   The physical interpretation of the model requires adequate
understanding of the excited states. In the present
section we shall consider the excited states as deviations from
the ground states determined in the previous section. The results
presented here for K=2 may be directly generalized to the case of 
arbitrary K \cite{Sztyren:2006}.

   For two reasons the problem will be treated in the linear context:\\
  (a) The nonlinearity is present in the equations through the term
\(|\psi|^2\) only, hence the excitations
which do not change the modulus \(|\psi|\) (we shall call them
phase excitations)
may be treated in the framework of linear equations exactly.\\
  (b) For the excitations changing the modulus of the order parameter,
one can
solve the problem in linear approximation, valid for weakly excited
states.\\
   From the microscopic point of view one speaks here on the
collective excitations of the Cooper pair condensate.\\

   Let us first consider the phase excitations. We substitute
the function
\equln{\psi_n=fe^{i\theta_n},\ {\rm where}\ f=const.
}{psiun}
into the equation
\equln{(\tilde\alpha+\beta|\psi|^2)\psi_n-\frac{1}{2}[
\gamma_1(\psi_{n+1}+\psi_{n-1})\psi_n+
\gamma_2(\psi_{n+1}+\psi_{n-1})\psi_n]=\epsilon\psi_n
.}{enun}
Then we obtain
\equln{\tilde\alpha+\beta|\psi|^2-\frac{1}{2}[
\gamma_1(e^{i(\theta_{n+1}-\theta_n)}+
  e^{i(\theta_{n-1}-\theta_n)})+\gamma_2(e^{i(\theta_{n+2}-\theta_n)}+
  e^{i(\theta_{n-2}-\theta_n)})]=\epsilon(\theta),
}{enun1a}
which should be independent of \(n\).
 In consequence, \(\theta_n=n(\theta_0+\theta) + {\rm const}\) and
\equln{\epsilon(\theta)=\tilde\alpha+\beta f^2 -
 [\gamma_1\cos{(\theta_0+\theta)}+\gamma_2\cos{2(\theta_0+\theta)}],}
{eph}
where \(\theta_0\) corresponds to the ground state determined in the
Section 5. The particular values \(\theta_0=0\) and
\(\theta_0={\pi}\) represent the uniform and alternating ground states,
respectively. The remaining values of \(\theta_0\) represent the
phase modulated ground states.\\

The expression
for the energy of a phase excitation has the form:
\equln{\epsilon(\theta)-\epsilon(0)=\gamma_1(\cos{\theta_0}
-\cos{({\theta}_{0}+\theta})+\gamma_2(\cos{2\theta_0}
-\cos{2({\theta}_{0}+\theta}))
,}{eph1}
where \(\epsilon(0)\) denotes the energy of ground state
\equln{\epsilon(0)=\hat\alpha(\theta_0)+\beta f^2
.}{eps0}

   Now we shall examine the modular excitations.
Let us consider once more the equation (\ref{enun}) but this
time we shall take into account that
\(\psi_n\ \rightarrow\ f+\delta\psi_n\), where
\(f\) denotes a uniform ground state. After linearization one obtains
equation of the same shape, with \(f^2\) in place of \(|\psi|^2\)
and \(\delta\psi\) in place of \(\psi\).
Taking excitation \(\delta\psi_n=ae^{inks}\) with arbitrary constant
amplitude \(a\) and wave vector \(k\) (it is sufficient to consider
excitations with \(k_x=0=k_y\)), after
some calculations like in
the case of phase modulated excitations, one obtains the excitation
energy corresponding to wave vector \(k\) (with respect to the ground
state \(k=0\))
\equln{\epsilon_1=\epsilon_k-\epsilon_0=\gamma_1(1-\cos{ks})
+\gamma_2(1-\cos{2ks})\ \ \ 
}{eph1a}
valid in the region of stability of uniform ground state:
\(\gamma_1>0,\ \gamma_1+4\gamma_2>0\).\\
   Similar procedure perfomed for the alternating ground state
delivers the formula
\equln{\epsilon_2=-\gamma_1(1-\cos{ks})
+\gamma_2(1-\cos{2ks})\ \ \ 
}{eph2}
in the region of stability of alterating ground state:
\(\gamma_1<0,\ -\gamma_1+4\gamma_2>0\).\\

   The case of phase-modulated ground state is more complicated
due to
the lack of invariance with respect to the time reversal. The result
in this case reads
\equln{\epsilon_3=
 \gamma_1(\cos{\theta_0}-\cos{(\theta_0+ks)})
 + \gamma_2(\cos 2{\theta_0}-\cos 2{(\theta_0+ks)})
}{eph3}
and is valid in the third region of stability of ground states:
\equln{\gamma_2<0,\ \ |\gamma_1|<4|\gamma_2|
}{south}
The quantity 
\(\theta_0\) in (\ref{eph3}) equals the value of the variable
\(\theta\) determined by (\ref{arccos}).
\section{Dispersion curves}
The calculation of the first and second derivatives of functions
\(\epsilon_1,\ \epsilon_2\) and \(\epsilon_3\)
describing excitation energy allows us to discuss
shapes of dispersion curves in several sectors of the plane
\((\gamma_1,\gamma_2)\) and to obtain the expressions for bandwidth W
and effective mass \(m_c\).
It is sufficient to plot the curves on the
interval \([0,{\pi}]\). The Fig. 1 contains examples of all 6 possible
shapes of dispersion curves.\\

 The first and second derivatives of excitation
energy are expressed by the same formulae for both the east and
nord-east sectors (the region of stability of the uniform state):
\equln{\epsilon_1'=s(\gamma_1+4\gamma_2\cos{ks})\sin{ks},\ \ 
\epsilon_1''=s^2\gamma_1\cos{ks}+4\gamma_2(2\cos^2ks-1)
.}{e1p}
The effective mass, expressed by \(\epsilon_1''(k=0)\), is then for both
E and NE sectors given by the same formula:
\equln{m_c=\frac{\hbar^2}{s^2}(\gamma_1+4\gamma_2)^{-1}
.}{m1}
However, the expressions for bandwidth are different.
In the E sector \(\epsilon_1'{\geq}0\), the dispersion curve is
monotonic, and the bandwidth W is expressed by the difference of values of
\(\epsilon_1\) at the ends of interval \([0,{\pi}]\):
\equln{W=2\gamma_1
.}{W1}
In the NE sector the dispersion curve reaches maximum at the point
\(ks=\theta\), where \(\theta\) is determined by (\ref{arccos}),
(\(\theta<{\pi}/2\) for NW sector and \(\theta>{\pi}/2\) for NE one).
The expression for bandwidth calculated as 
\(\epsilon_1(\theta)-\epsilon_1(0)\) has the form
\equln{W=\gamma_1+2\gamma_2+\frac{\gamma_1^2}{8\gamma_2}.
}{W2}
\indent
   In the region of stability of the alternate state, i.e. in the
nord-western
and western sectors, the formulae for the first and second derivatives
of excitation energy \(\epsilon_2\) have the form (\ref{e1p}) with 
\(\gamma_1\) replaced by \(-\gamma_1\). Hence the
substitution of \(|\gamma_1|\) instead of \(\gamma_1\) into (\ref{W1})
gives bandwidth for western sector; the same substitution into (\ref{W2})
leads
to the bandwidth for nord-western sector. The expression for effective
mass for both NW and W sectors may be
obtained from (\ref{m1}) by the same substitution again and has the form
\equln{m_c=\frac{\hbar^2}{s^2}(|\gamma_1|+4\gamma_2)^{-1}
.}{mef}
\indent
   For the region of stability of the phase-modulated foundamental states,
i.e. for both southern sectors, we have
\equln{\epsilon_3'=s[\gamma_1+4\gamma_2\cos(\theta+ks)]
\sin(\theta+ks)
}{e3p}
and
\equln{\epsilon_3''=s^2[\gamma_1\cos(\theta+ks)
+4\gamma_2(2\cos^2(\theta+ks)-1)
}{e3b}
with \(\theta\) expressed by (\ref{arccos}). In the interval \([0,{\pi}]\)
 \(\epsilon_3'=0\) if \(\theta+ks=0\) or \(\theta+ks={\pi}\), or else
\(\cos(\theta+ks)=cos(\theta)\). That implies that \(\epsilon_3\) has
minimum at \(ks=\theta\) (\(\theta>{\pi}/2\) for SW sector and \(\theta<{\pi}/2\)
for SE one).\\

   The expression for bandwidth in SW sector is equal to the difference\\
\(\epsilon_3(\theta)-\epsilon_3(0)\) and in SE sector
\(\epsilon_3({\pi})-\epsilon_3(\theta)\). The common
formula for the bandwidth has the form
\equln{
W=|\gamma_1|+2|\gamma_2|+\frac{\gamma_1^2}{8|\gamma_2|}.
}{bandNS}
while the effective mass in both southern sectors is given by
\equln{m_c= \frac{\hbar^2}{s^2}
            \frac{4|\gamma_2|}{16\gamma_2^2-\gamma_1^2}.
}{ms}
This completes the discussion of bandwidths and effective masses for
all the types of ground states.
\thebibliography{12}
\bibliography{}
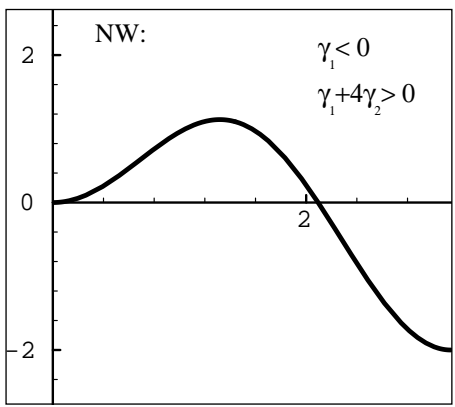
\end{document}